\documentclass[]{spie}  

\usepackage{amsmath,amsfonts,amssymb}
\usepackage{graphicx}
\usepackage[colorlinks=true, allcolors=blue]{hyperref}

\newcommand{\nii}{\textsc{[Nii]}$\lambda$6583}
\newcommand{\oiii}{\textsc{[Oiii]}$\lambda$5007}

\title{Realizing the potential of the Dragonfly Spectral Line Mapper: Calibration methods and on-sky performance}

\author[a]{Deborah M. Lokhorst}
\author[b,c]{Seery Chen}
\author[d]{Imad Pasha}
\author[a]{Victoria Purcell}
\author[d]{William P. Bowman}
\author[b,c]{Qing Liu}
\author[d]{Zili Shen}
\author[a]{Aidan MacNichol}
\author[e]{Evgeni I. Malakhov}
\author[b,c]{Roberto G. Abraham}
\author[d]{Pieter van Dokkum}

\affil[a]{NRC Herzberg Astronomy \& Astrophysics Research Centre,
5071 West Saanich Road, 
Victoria, BC V9E2E7, Canada}
\affil[b]{David A. Dunlap Department of Astronomy \& Astrophysics,
University of Toronto,
50 St. George Street, 
Toronto, ON M5S3H4, Canada}
\affil[c]{Dunlap Institute,
University of Toronto,
50 St. George Street, 
Toronto, ON M5S3H4, Canada}
\affil[d]{Department of Astronomy,
Yale University,
52 Hillhouse Ave., New Haven, CT 06511, USA}
\affil[e]{New Mexico Skies, Inc., 
9 Contentment Crest, Mayhill, NM 88339, USA}

\authorinfo{Send correspondence to D.M.L.: deborah.lokhorst@nrc-cnrc.gc.ca}

\begin{document}
\maketitle

\begin{abstract}
The Dragonfly Spectral Line Mapper is an innovative all-refracting telescope designed to carry out ultra-low surface brightness wide-field mapping of visible wavelength line emission. Equipped with ultranarrowband (0.8 nm bandwidth) filters mounted in Dragonfly Filter-Tilter instrumentation, the Dragonfly Spectral Line Mapper maps H$\alpha$, \nii, and \oiii~line emission produced by structures with sizes ranging from $\sim$1 to 1000 kpc in the local Universe. These spatial scales encompass that of the exceedingly diffuse and faintly radiating circumgalactic medium, which is singularly difficult to detect with conventional mirror-based telescope instrumentation. Extremely careful control of systematics is required to directly image these large scale structures, necessitating high fidelity sky background subtraction, wavelength calibration, and specialized flat-fielding methods. 
In this paper, we discuss the on-sky performance of the Dragonfly Spectral Line Mapper with these methods in place. 
\end{abstract}



\section{INTRODUCTION}
\label{sec:intro}  

The Dragonfly Spectral Line Mapper\cite{chen22} (DSLM) is a telescope built to image the circumgalactic medium of galaxies in the local Universe. The circumgalactic medium, along with the intergalactic medium, trace the `cosmic web' of dark matter, which dictates where large scale structures, such as galaxies, form. Gas is predicted to flow along the filaments of the cosmic web (the intergalactic medium), transition into the atmospheres of galaxies (the circumgalactic medium), and eventually flow into galaxies. The method by which galaxies accrete gas from the circumgalactic medium is still unknown at low redshifts, with recent simulations suggesting that cold streams play a significant role\cite{loch20}. The circumgalactic medium is diffuse and faintly emitting, and at low redshifts, the bright ultraviolent emission lines are not accessible by ground-based facilities. In addition, the large spatial scales of the low redshift circumgalactic medium require large fields of view\cite{lokh22b}, making the observation of the circumgalactic medium more suited to narrowband imaging rather than integral field unit spectroscopy. Conventional narrowband imaging is limited in spectral resolution and surface brightness, though, and thus has difficulty reaching the level that is required to detect the circumgalactic medium ($\sim10^{-20}$~erg~cm$^{-2}$~s$^{-1}$~arcsec$^{-2}$)\cite{lokh19}. DSLM combines the extremely low surface brightness sensitivity of the Dragonfly Telephoto Array\cite{abra14} with innovative instrumentation (the Dragonfly Filter-Tilter\cite{lokh20}) to reach unprecedented surface brightness sensitivity to visible wavelength line emission, enabling it to directly detect the circumgalactic medium of local galaxies. This ability was demonstrated in imaging from the pathfinder DSLM\cite{lokh22a}.

DSLM is a fully refracting mosaic-design telescope that is based on the instrumentation of the Dragonfly Telephoto Array. The Dragonfly Telephoto Array is a specialized low surface brightness imaging telescope built to map stellar continuum light. DSLM is an expansion to the Dragonfly Telephoto Array, utilizing the same instrumentation with the addition of Dragonfly Filter-Tilters, which hold an ultranarrow-bandpass filter in front of the optics of each unit of the telescope array.
DSLM comprises 120 commercial telephoto lenses, each equipped with commercial off-the-shelf focusers and cameras, as well as Dragonfly Filter-Tilters\cite{lokh20}.
The details of the design of DSLM are described in Refs.~\citenum{lokh20,lokh22b,chen22}, while we provide a brief overview here.

The main instrumental component of DSLM is the Dragonfly Filter-Tilter which holds a large format ultranarrowband filter. The ultranarrow filter bandpasses enable DSLM to map ultra-low surface brightness visible line wavelength emission. A Dragonfly Filter-Tilter is positioned in front of each lens on the array, which allows the filter bandpasses to be narrowed beyond conventional limits, as it is at the entrance pupil of the optical system and within a collimated beam. DSLM employed two types of ultranarrowband filters for scientific imaging which have bandpasses with a FWHM of 0.8 nm and central wavelengths of 664.7nm and 507.1nm. The longer wavelength filter is intended to detect H$\alpha$ and \nii~ line emission separately, which the shorter wavelength filter is intended for \oiii~ line emission.

As of November 2023, there are 100 science filters and 10 background filters on DSLM (with an additional 10 filters to be installed at a later date). As the name suggests, the filter-tilters are able to rotate the filters, which smoothly shifts the central wavelength of the filter bandpass. With a rotation of up to 20 deg, the bandpass shifts by $\sim$8 nm. This allows the filters to cover the same cosmological volume as a typical $\sim10$ nm bandpass narrowband filter, but with the benefit of reducing the sky background continuum by over an order of magnitude, and the ability to completely remove strong sky lines. The science filters tilt through a redshift range from z=0 to z=0.01.  Each unit of DSLM has a field of view (FOV) of $1.5^{\circ}\times1.9^{\circ}$ and pixel scale of $2.45''$, which yields angular resolution varying from $\sim10 - 200$ pc over a FOV of $\sim30 - 1000$ kpc through the target redshift range.

To carry out successful imaging of low surface brightness structures on large scales, the science images need to be calibrated to less than a percent of the background level in the images. 
The circumgalactic media of galaxies is conventionally defined to extend out to the virial radii of the galaxy, which results in an extremely large angular size for local galaxies (e.g., a Milky Way mass galaxy with a total mass of $\sim 10^{12}$ M$_{\odot}$ has a virial radius of $\sim200$ kpc\cite{dehn06}). The circumgalactic media for galaxies with redshifts z$<0.01$ encompass the majority of the field of view of the detectors, requiring extremely careful calibration on large scales.
For the Dragonfly Telephoto Array, this is achieved through careful flat-fielding and background subtraction modeling. We aim to achieve similar limits in flat-fielding and background subtraction. The narrowness of the science filter bandpasses are a barrier to the methods used in the Dragonfly Telephoto Array, requiring additional processes to complete the required calibration. In Section~\ref{sec:calmethods}, we describe two of these methods in particular. 
In Section~\ref{sec:flats} a new method for the collection of flat-field frames based on a large electroluminscent panel is described.
In Section~\ref{sec:ftcal}, we describe the latest filter-tilter calibration methods employed on DSLM.
Section~\ref{sec:onsky} describes the on-sky performance of the DSLM, previewing images from each type of filter and emission line, along with a discussion of the effect of the field of view on the science filter bandpasses.

\section{CALIBRATION METHODS}
\label{sec:calmethods}

While DSLM was being constructed from March 2022 to November 2023, new calibration methods for the telescope array were developed, implemented, and tested. There are two main calibration methods that needed to be implemented. The first is a flat-field image collection method, the development of which is described in Section~\ref{sec:flats}. Flat-field images are required to correct for differences in the transmittance of light across the field of view and pixel-level sensor quantum efficiency variations. A myriad number of effects are accounted for by flats, such as dust on the optics, detector sensitivity variations, and angular dependence in the field illumination. Flat-field calibration introduces a multiplicative error if done improperly.
The second method is a filter-tilter calibration method, which is described in Section~\ref{sec:ftcal}. The filter-tilter calibration is required to ensure that the bandpass of the filters has been correctly shifted to be centred on the wavelength of interest. If the calibration is not carried out successfully, the throughput of the filters will be strongly degraded, resulting in poor sensitivity. Furthermore, our imposed model of the filter central wavelength across the field of view will be incorrect, resulting in poor flux calibration.

\begin{figure}[t]
  \begin{center}
  \begin{tabular}{c} 
  \includegraphics[width=0.9\linewidth]{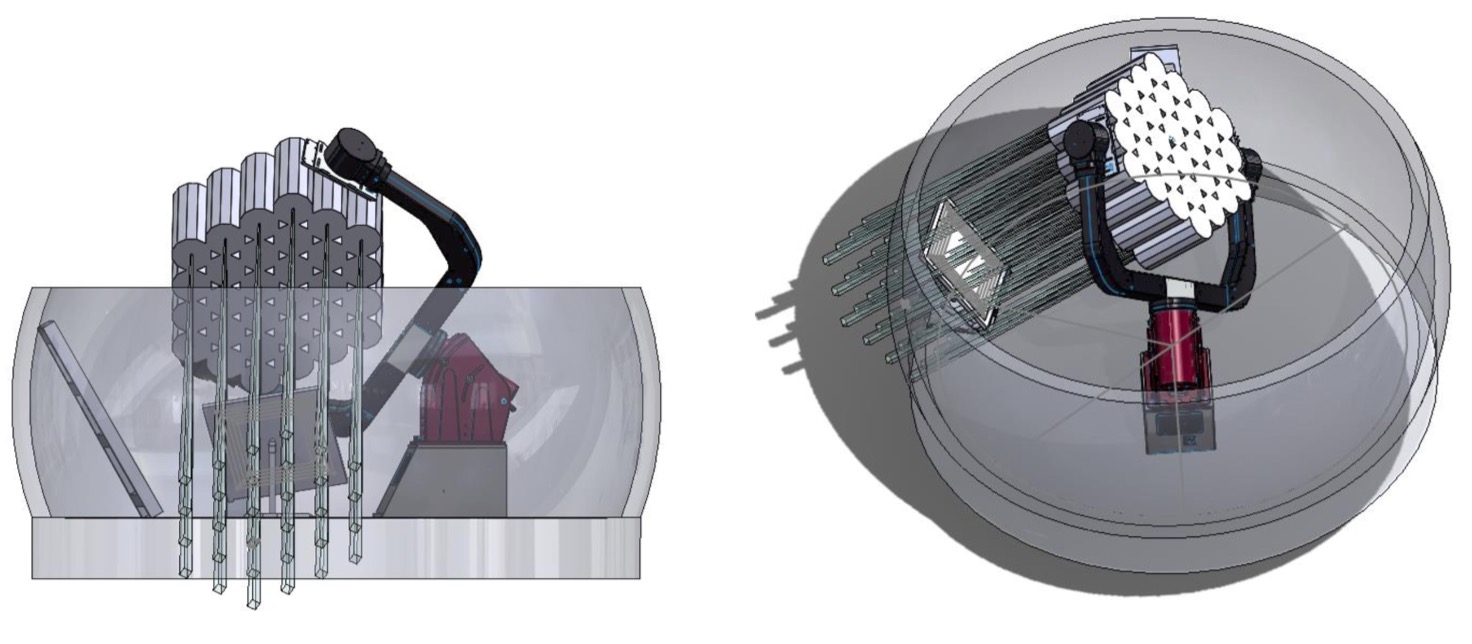}
    \end{tabular}
  \end{center}
  \caption{ \label{fig:panelCAD} 
  An electroluminescent panel is installed on the floor of the Dragonfly Spectral Line Mapper dome for flat field image collection. In the shown CAD model, each unit's FOV is projected to demonstrate their intersection with the panel. Multiple telescope pointings are required for all the units to view the panel. Flat field image collection is carried out in groups of up to four units at a time.
}
\end{figure} 

\subsection{Flat Fielding Panel Implementation}
\label{sec:flats}

The addition of ultranarrowband filters introduces several complexities to the basic method of collecting flat field images during twilight and dawn (which is the method used successfully by the Dragonfly Telephoto Array). First, the narrowness of the filter bandpasses can result in strong variation in the sky background (see Section~\ref{sec:onsky}). This occurs when the filter tilts used for science data collection contain a strong sky line(s) in addition to the targeted emission line. When collecting flat-field images at these tilts (to ensure that centroid locations and light paths match with the science data)\cite{chen22}, the flat field images contain features introduced by sky lines, which makes them unsuitable for use in flat field calibration. In addition, the narrow filter bandwidths require longer exposure times, which reduces the number of flat-field images that can be collected each night. This time crunch is exacerbated by the need to collect flat-field images at each tilt of the filter used for science imaging. Even without the effect of the sky line variation within the fields of view of the detectors, it would be difficult to collect sufficient flat-field calibration frames during twilight and dawn.

The pathfinder DSLM\cite{lokh22b}, which was a 3-lens version of the full DSLM that was on-sky from the spring of 2019 to the fall of 2021, used Aniltak Flip-Flats to collect flat field images. These were electroluminescent panels that were mounted on each unit of the array to mitigate the time constraint on collecting flat-field frames (note that the pathfinder DSLM filters were not as narrow as the filters used for the full DSLM, with bandwidths of 3 nm, so the sky line variation was not a significant issue). The Flip-Flats are not possible to implement for DSLM due to the tight physical positioning of all the units. The motors and panels of the Flip-Flats would block the apertures of adjacent units when they are positioned in the perpendicular `open' position for nightly observing. The additional weight from Flip-Flats on each unit would also interfere with the mount slewing capability. 

In consideration of these issues, we developed a new method for collecting flat field images, which utilizes a large (24 inch by 24 inch) electroluminescent panel (an Alnitak Flat-Man XL-2) mounted on the dome floor (Fig~\ref{fig:panelCAD}). In this location, the panel has no interference with the field of view of DSLM during nightly observations. Due to the compact array, which is less than 1 m long along the optical axis, the telescope can slew 180 deg around its polar axis (i.e., Right Ascension). This allows the telescope to point at the dome floor, or in this case, directly at a panel positioned in a mount on the dome floor. We demonstrate the location of the panel in a CAD shown in Fig~\ref{fig:panelCAD}. Currently, there is only one panel installed in one of the Dragonfly Spectral Line Mapper domes. Additional panels are planned to be installed in the other three domes.

 \begin{figure}[t]
  \begin{center}
  \begin{tabular}{c} 
  \includegraphics[width=0.35\linewidth]{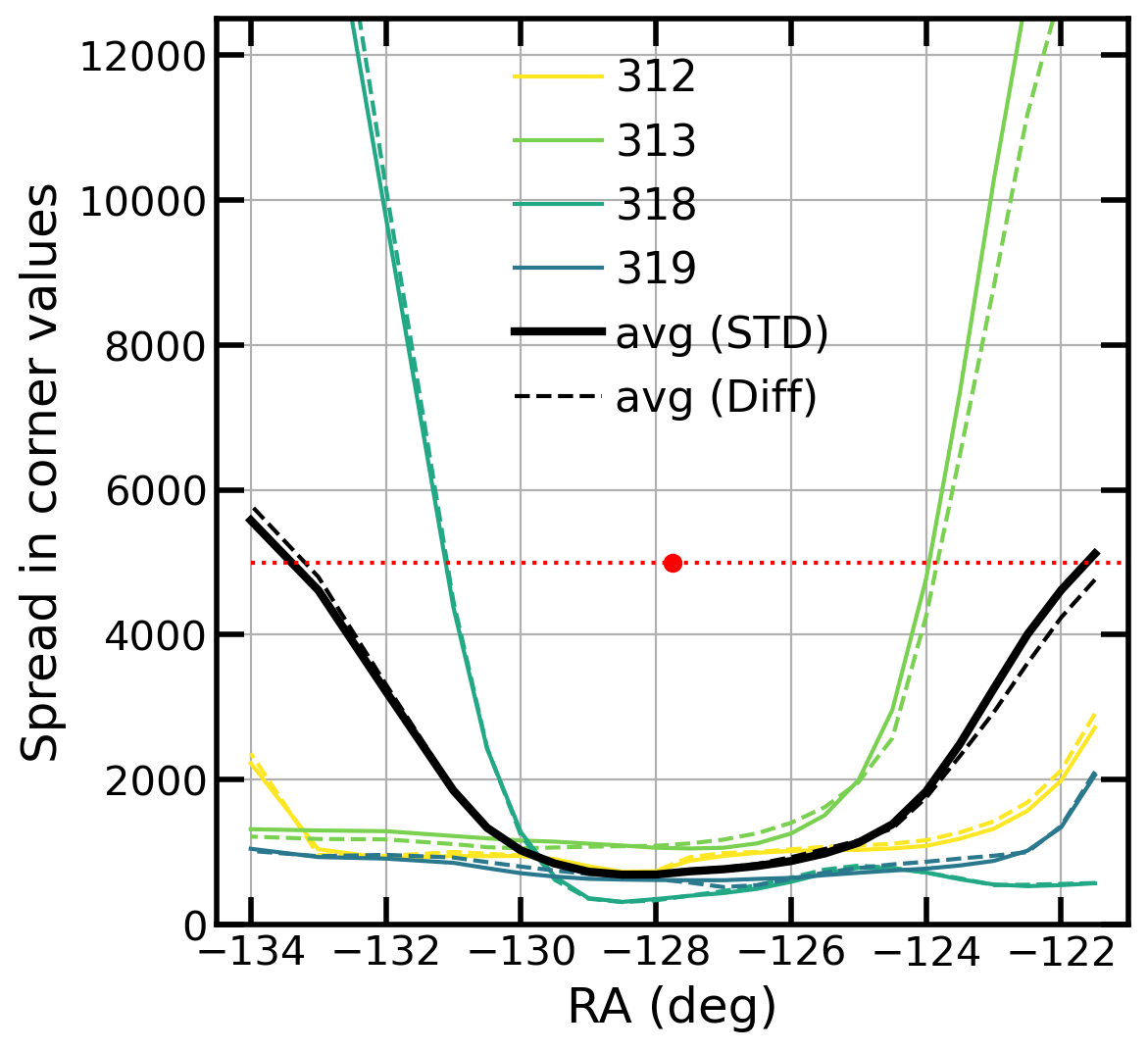}
  \includegraphics[width=0.345\linewidth]{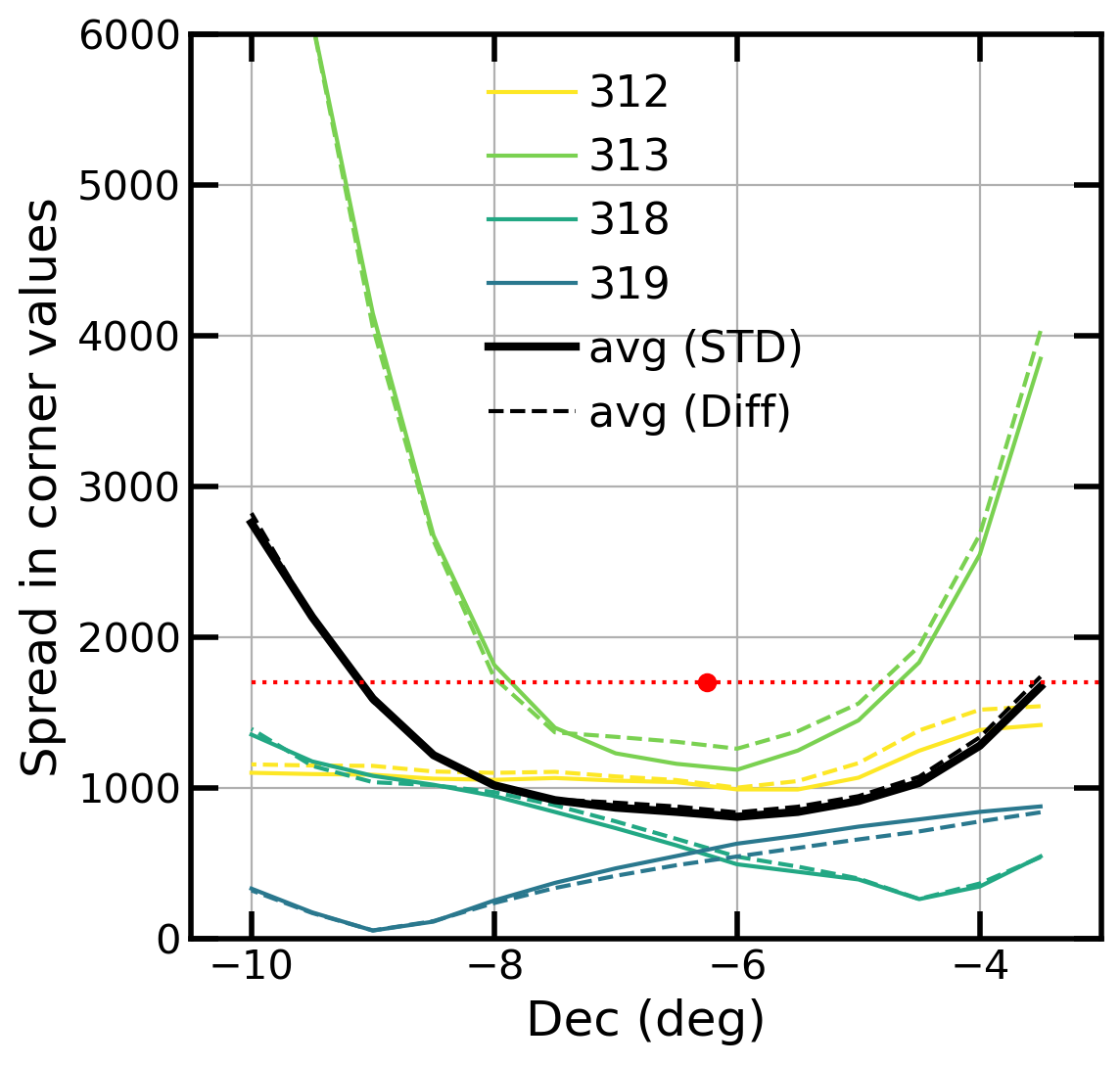}
    \end{tabular}
  \end{center}
  \caption{ \label{fig:panelcalib} 
  Demonstration of the calibration of the telescope pointing onto the flat panel. After centering the telescope pointing for one group of lenses (the middle group in this case, which contains four units), we slew over the flat panel over an about 10 deg span in both horizontal (RA) and Declination axes. We collect flats from the group of units at each pointing, then use the ramp of the brightness across each flat to determine the best pointing position (red dot).
}
\end{figure}

The close placement and relatively small size of the electroluminescent panel requires multiple pointings of the telescope for all units to gather flat-field frames by imaging the panel. Optimizing for even illumination for each FOV for each individual unit, we use nine pointings to collect a full set of flat-field frames. The number of lenses that could collect flats at the same time was maximized while avoiding a four inch rim at the edge of the panel that did not have constant illumination. The resulting groups ranged from one to four units that could view the panel at the same time and collect flats.
Fig~\ref{fig:panelCAD} includes a simplified CAD model of the DSLM array, shown as a honeycomb-type structure attached within the forks of the telescope mount (Software Bisque Taurus 700). The projected FOVs of each unit are shown from the array, demonstrating the overlap of the FOVs with the panel.

The telescope required calibration to correctly point to the panel after the panel was mounted in the dome. We implemented a simple yet effective method to center the telescope pointing onto the panel. This pointing calibration is carried out using the middle four lenses in the honeycomb array. The telescope slewed to point that grouping of lenses at the panel, then the telescope was varied in both Dec and RA over a range of 10 deg in each direction. Flats were collected from the panel at 0.5 deg increments. We then performed an analysis on the flats to determine the steepness of a brightness gradient across each flat, using that gradient to determine the midpoint for both Dec and RA individually. This yielded a correction to the estimated pointing for that group of lenses. This correction was applied to the pointings for the other lens groupings using a simple scaling relationship. An example of the flat analysis results for the calibration of the telescope pointing onto the panel is shown in Fig~\ref{fig:panelcalib}.


\subsection{Dragonfly Filter-Tilter Calibration}
\label{sec:ftcal}

As described in Refs~\citenum{lokh20,lokh22b}, the filter bandpasses can be shifted by varying the angle of incidence of light upon the filter. The relationship between the angle of incidence and the central wavelength of the filter bandpass is a smoothly varying, monotonic function which is well defined by the effective index of refraction of the filter. 

\begin{figure}[t]
  \begin{center}
  \begin{tabular}{c} 
  \includegraphics[width=0.48\linewidth]{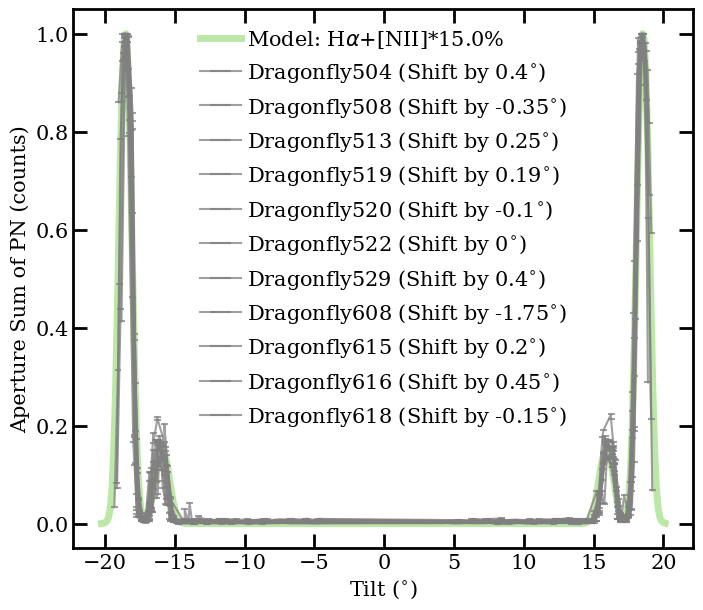}
  \includegraphics[width=0.48\linewidth]{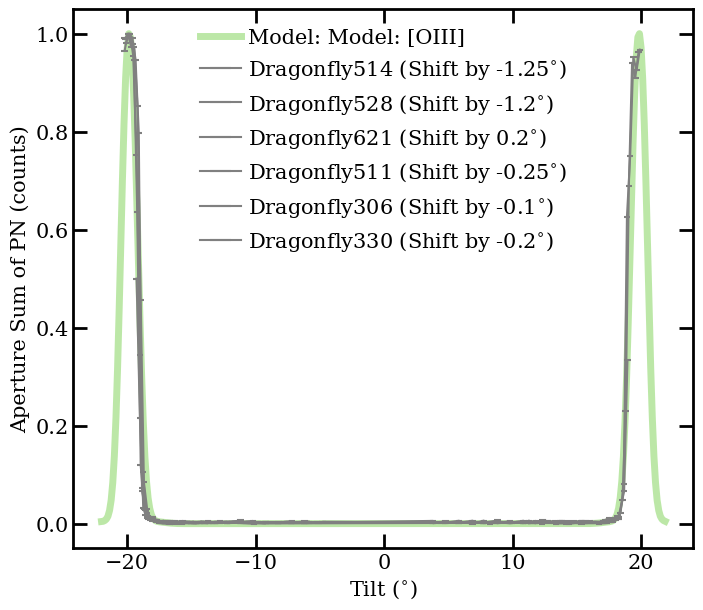}
    \end{tabular}
  \end{center}
  \caption{ \label{fig:ftcal} 
  Demonstration of the calibration of the filter-tilters, which is carried out by imaging a planetary nebula (PN). The filters are rotated through -20 to 20 deg, and images of the PN are taken at 0.25 deg increments. The flux of the PN is measured from the images and compared to a model of the flux from the PN. The top and bottom panels show an example of the calibration data from several units for the H$\alpha$ and \textsc{[Oiii]} filters, respectively. See Ref. 4 for more details.
}
\end{figure}

The filter-tilters are calibrated through on-sky measurements of a planetary nebula. NGC 6543 (the Cat's Eye Nebula) was used for the majority of the filter-tilter calibration. The filter-tilter calibration process followed the description in Ref.~\citenum{lokh22b}, where the telescope was pointed at a planetary nebula and images were collected at a series of filter angles. The flux of the planetary nebula in line emission is modeled by assuming a gaussian line emission profile and integrating over the filter bandpasses at each filter tilt. For NGC 6543, a relative line brightness of $f_{[NII]} = 0.15 \times f_{H\alpha}$ was implemented in the flux model. Note that the DSLM science filters with central wavelengths of 664.7 nm are designed to image both H$\alpha$ and \nii~ line emission. For the same target, these two lines can be independently measured by tilting the filter to separate angles. For example, for NGC 6543, the H$\alpha$ and \textsc{[Nii]} emission lines are detected at tilts of 18.3 nm and 15.9 nm, respectively. The results of the filter-tilter calibration data collection and comparison with the modeled flux are shown in Fig~\ref{fig:ftcal}. 

This process was carried out iteratively to avoid running into the hard rotation limits for the filter-tilter, which is restricted to rotate between approximately -20 deg to 20 deg. The central wavelengths of the science filters reach rest line wavelength for H$\alpha$ and \textsc{[Oiii]} at tilts of 18.3 deg and 19.6 deg, respectively.  Since the range of allowed angles for the filter-tilter rotation is constrained to an absolute maximum of 20 deg, we begin the calibration with a range up to 17.5 deg, allowing for large adjustments to avoid the filter-tilters running repeatedly into the hard limits and potentially getting stuck.
The filter-tilter calibration was also affected by hardware limitations, notably that there were a significant number of units that could not focus. This affected the accuracy of the filter-tilter calibration. For units that were in focus and had good image quality during the filter-tilter calibration data collection, the filter-tilter calibration was successful to $\pm 0.1$ deg accuracy.  A tilting precision of 0.4 deg is required for the narrow bandwidths to maximize the bandpass across the FOV\cite{lokh20}. The filter-tilter motors have a rotational accuracy of 0.1 deg\cite{chen22}. Accounting for these sources of error, the tilting accuracy meets the requirements for on-band data collection. 

This estimation of required filter-tilter accuracy does not take into account the potential offset in the angular location of the target within the field of view (such as if the target is an extended source). This introduces an additional uncertainty on how well the bandpass is centred on the line wavelength of the target of interest at its position in the field of view. With this in mind, we aim to improve the accuracy of our filter-tilter calibration in the future. The uncertainty in the filter-tilter calibration is largely driven by the high filter tilt required to shift the bandpass to the line emission of the planetary nebulae. While the line emission from extragalactic targets could be used (such as from a bright starburst galaxy), the emission is significantly fainter than the emission from planetary nebulae, which introduces additional uncertainty into the measurement.
An alternative method that we are currently considering is to incorporate a laser-lit target with a wavelength corresponding to $\sim10$ deg tilt to increase the brightness of the calibration source and the ease of filter-tilter calibration. A laser-lit target was used for the pathfinder DSLM filter-tilter calibration, but was difficult to implement due to the target being positioned near the dome. The nearness of the target required individual pointings of the telescope for every single lens. This increased the complexity and time required for filter-tilter calibration to being nearly unattainable for the full DSLM (even with a pointing routine to slew the telescope, the uncertainty in pointing and required time to collect images individually for each unit is prohibitive). An alternative option we are considering is to mount or fly via drone a target at a significant distance ($\gtrapprox 100$ m) from the telescopes. This would enable the target to show up in a single pointing, but presents challenges in either the required altitude and location of ground-based mounting, or the stability of the target during flight. Another potential calibration source could be a strong sky line, which are prevalent within the wavelength range covered by the H$\alpha$/\textsc{[Nii]} filter (see Section~\ref{sec:onsky} for a further discussion of sky lines).


\begin{figure}[t]
  \begin{center}
  \begin{tabular}{c} 
  \includegraphics[width=0.97\linewidth]{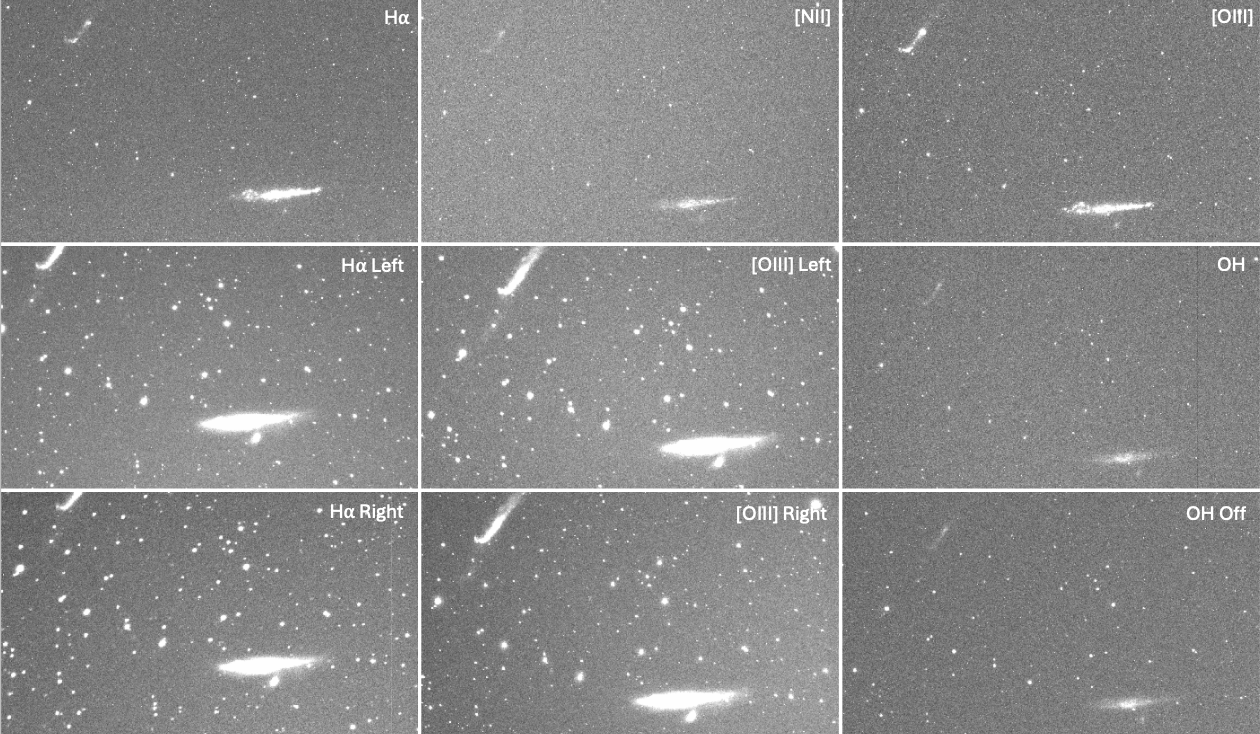}\\
    \end{tabular}
  \end{center}
  \caption{ \label{fig:imcutouts} 
    Cutouts from images from each type of filter and line emission  are displayed. Example images from the science filters are shown across the top, demonstrating the difference between the H$\alpha$, \textsc{[Nii]} and \textsc{[Oiii]}  line emission from the NGC 4631 and NGC  4656 galaxies.
}
\end{figure} 

\section{ON SKY PERFORMANCE}
\label{sec:onsky}

The Dragonfly Spectral Line Mapper was completed in November 2023, and an initial survey of the circumgalactic media of galaxies in the local Universe is currently underway. Examples from the initial data collection on the NGC 4631 and NGC 4656 galaxies are shown in Fig~\ref{fig:imcutouts}. The H$\alpha$/\textsc{[Nii]} and \textsc{[Oiii]} filters are used to collect emission line data, while the OH and OH Off filters are used to subtract the sky background and the four off-band filters (H$\alpha$ Left, H$\alpha$ Right, \textsc{[Oiii]} Left, and \textsc{[Oiii]} Right) are used to subtract continuum light (see Ref.~\citenum{chen22} for more details on the filter bandpasses and design). The H$\alpha$ and \textsc{[Nii]} images were taken with the same filter at different tilts to select the wavelength of the line emission of interest (657.6 nm and 659.6 nm, respectively). The difference in morphology and brightness between the three emission lines is apparent; the H$\alpha$ and \textsc{[Oiii]} emission from the galaxies is similar in brightness, while the \textsc{[Nii]} emission is noticeably fainter. 
The sky background in the \textsc{[Nii]} data is also notably brighter than the background in the H$\alpha$ and \textsc{[Oiii]} imaging, which is due to the moon being present during the \textsc{[Nii]} imaging. 

\subsection{Bandpass variation across the field of view}

\begin{figure}[ht]
  \begin{center}
  \begin{tabular}{c} 
  \includegraphics[width=0.97\linewidth]{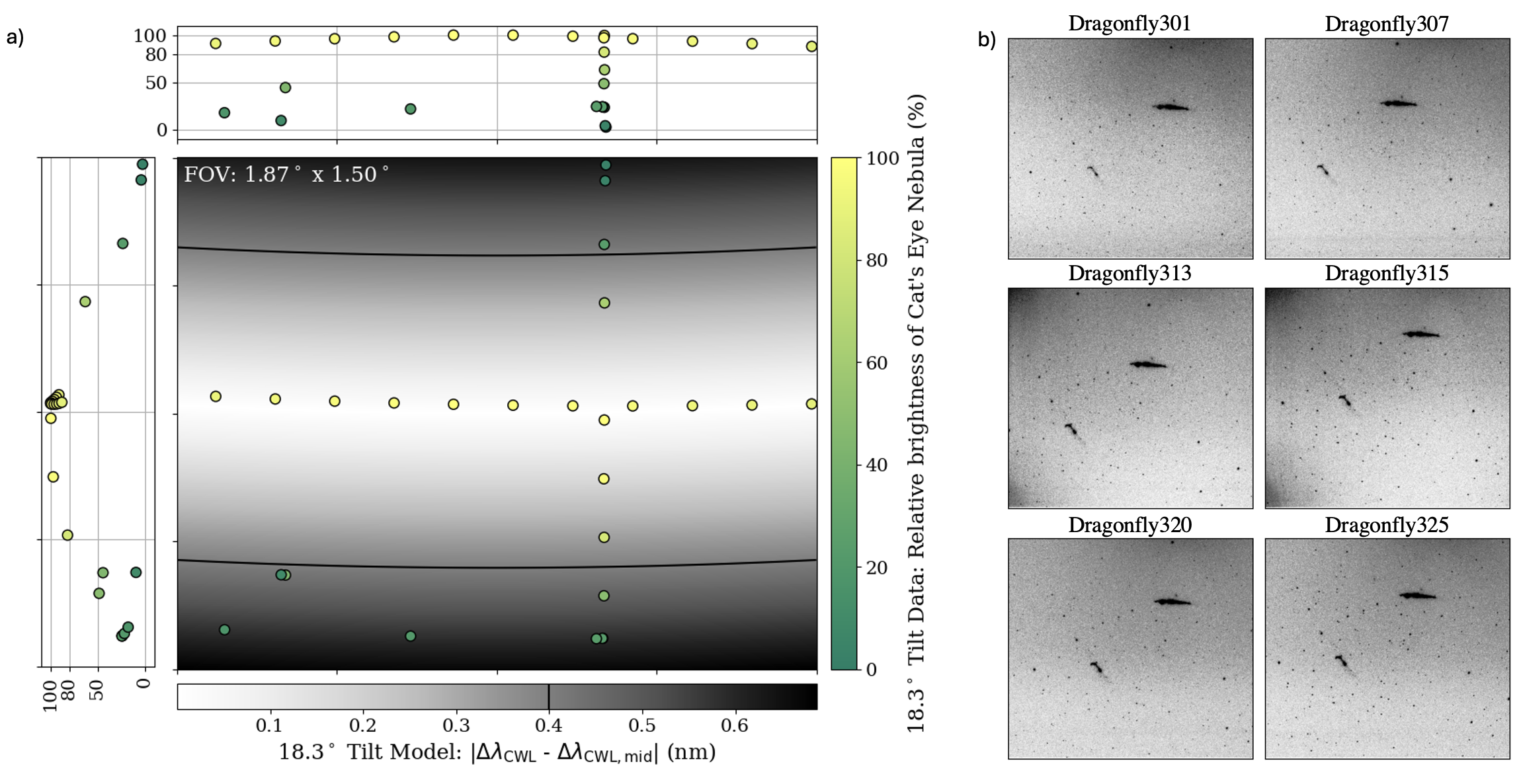}

    \end{tabular}
  \end{center}
  \caption{ \label{fig:model} 
    The effective filter bandpass strongly  depends on the  filter tilt and position in the field of view (FOV). \emph{Panel a):} The greyscale image is a model of the shift in filter CWL across the field of view. The black lines delineate the limits of the 0.8nm bandpass within the FOV. This is verified with planetary nebula flux measurements across the FOV (overplotted). 
    \emph{Panel b):} The images demonstrate this effect via a sky line which is near in wavelength to the H$\alpha$ line emission from NGC 4631. This sky line causes a strong vertical gradient in the sky background brightness.
}
\end{figure} 


We use the initial survey data to evaluate the performance of the filter-tilters and filters. In particular, we investigate the throughput of the filter as a function of position in the field of view. As discussed in Ref~\citenum{lokh20}, the angular shift across the field of view introduces a secondary `tilt' into the system, which is added in quadrature with the tilt introduced by rotating the filter. Understanding and verifying the additional component of the angle of the filter with respect to the optical axis is important for two main calibration processes: 1) modeling and removing the sky background to the required level and 2) carrying out accurate flux calibration across the field of view. In this section we aim to confirm our current model of the filter throughput across the field of view.

This shift is modeled for several filter tilts in Ref~\citenum{lokh20}; we apply the same model for the current DSLM science filters here. In panel a) of Fig~\ref{fig:model}, we show a model of the shift in the central wavelength of the H$\alpha$/\textsc{[Nii]} filter when the filter is tilted at an 18.3 deg angle. This figure demonstrates that the central wavelength changes by more than $\pm$0.6 nm across the FOV. The black lines on the model delineate the limits of the 0.8nm bandpass within the field of view; outside of these black lines, the bandpass has shifted to no longer encompass the wavelength of the line emission of the target (with a transmission greater than 50\%). In other words, for the ultranarrowband filters, the field of view which collects data at the wavelength of interest (the `effective' field of view) is a function of the tilt of the filter. This is noticeable in the field of view collected by DSLM due to the large size of the detector, which is $\sim 1.5^{\circ} \times 2^{\circ}$ in size. The effect comes into play when the filter is tilted to $\gtrapprox 11$ deg. 
At filter tilts less than $\approx 11$ deg, the shift of the central wavelength across the detector field of view is less than the filter bandpass so the effective field of view is unchanged from the detector field of view. 
At filter tilts approaching 20 deg, the field of view where the line emission from the target of interest falls within the filter bandpass becomes as low as $\approx$ 56\% of the total detector field of view. In panel a) of Fig~\ref{fig:model}, the effective field of view is $\approx$ 60\% of the total field of view. 


We confirmed the model of the filter throughput across the field of view by collecting data from a planetary nebula across the field of view. The planetary nebula NGC 6543 was imaged at positions across the field of view varying in both declination and right ascension with the H$\alpha$/\textsc{[Nii]} filter.
The flux of NGC 6543 was calculated in each image and the relative flux is plotted along the left and top sides of panel a) in Fig~\ref{fig:model}. The flux swiftly drops off around the rotation axis of the filter shown in the left plot, which is aligned with declination and the short axis of the detector. The results from the on sky flux measurement of NGC 6543 are shifted slightly from the model, which is most likely due to the tilt not being exactly centered on the emission line of NGC 6543. With this consideration, the flux profile measured from NGC 6543 fits the model well.
Along the middle axis of the detector (constant right ascension), the flux measured from NGC 6543 varies by $<10\%$, which is expected since the only tilt along the right ascension axis is that introduced by the field of view itself (e.g., $\pm 0.94$ deg relative to the center of the detector).

The changing central wavelength of the filter bandpass across the field of view can also be seen in the H$\alpha$ science images of the NGC 4631 galaxy, which are shown in panel b) of Fig~\ref{fig:model}. The H$\alpha$ emission line wavelength of the NGC 4631 galaxy is near the wavelength of a strong OH sky line, which results in a sky background feature that changes in strength vertically across the field of view due to the shift in the filter bandpass. As can be seen in this imaging, careful understanding of the varying bandpass across the field of view is required to properly understand and model the sky background, especially in the cases where strong sky lines are present within the filter bandpass.

Understanding the change in the wavelength-dependent throughput of the filter bandpass across the field of view is the first step towards accurate sky background removal and flux calibration. 
Utilizing the model shown here for the varying bandpass across the field of view, we plan to model and subtract the sky background from the data, then create a model for the flux calibration across the field of view. This allows us to include the data that is outside of the effective field of view, as this data is still valid science data, just at a different wavelength.
In fact, by collecting data at `microtilts', where the bandpass is not shifted completely but instead stepped slowly across the field of view, DSLM is effectively a wide-field integral field unit. A cube of spectral information over the wavelength range encompassed by tilting the filter ($\sim$8 nm) can be collected.
We note that the uncertainty introduced by the filter-tilter calibration is a systematic uncertainty which could be addressed by using the `microtilting' method. With microtilting data, one could determine a correct pointing for that specific target by examining the emission line flux as a function of tilt in a similar method to the filter-tilter calibration discussed above. 





\section{Summary}

DSLM is a novel telescope built entirely from commercial off-the-shelf components, which is designed to carry out visible line wavelength mapping of the circumgalactic medium of galaxies in the local Universe. DSLM incorporates a novel method for carrying out ultranarrowband imaging, whereby ultranarrow-bandpass filters are mounted in front of the optics. In this location, the filters are within a collimated beam of light and the filters can be rotated, which smoothly shifts the central wavelength of their filter bandpass. DSLM employs two types of ultranarrowband filters: 1) filters with a central wavelength of 664.7 nm which are used to detect H$\alpha$ and \nii~ line emission and 2) filters with a central wavelength of 507.1 nm which are used to detect \oiii~ line emission. The $\sim 1.5^{\circ} \times 2^{\circ}$ field of view of DSLM is well suited to large spatial structures, matching the angular size of the circumgalactic medium of low redshift ($z < 0.01$) galaxies. In order to robustly measure features on these large angular scales, careful calibration of the hardware and data is required.

We discuss here several implications and complexities that arise from the use of ultranarrow-bandpass filters. We begin with a discussion of a new flat-fielding data collection technique, which is needed due to the prevalence of sky lines which have a strong variation across the DSLM field of view due to the extreme narrowness of the filters and the filter tilting method. We have implemented a flat-fielding panel mounted in the dome for flat-field collection.

We discuss the filter-tilter calibration method, which is needed to accurately tilt the filters. The filter-tilter calibration method currently uses an astrophysical target to calibrate to the correct wavelength and tilt. This is difficult to carry out with the ultranarrowband filters as the brightest astrophysical targets with H$\alpha$ and \textsc{[Oiii]} line emission are within the Milky Way galaxy, which is at the extreme tilting range of the filter-tilters. This introduces uncertainty in fitting the data due to incomplete collection across the modeled flux curve. Extragalactic targets can be used, but those have large uncertainties due to the faintness of the targets. We discuss alternative methods which we intend to implement going forward.

Finally, we present a preview of the data collected from the science and off-band filters of the DSLM. We follow this with a discussion of the variation of the filter central wavelength across the field of view. We confirm the modeled variation with measurements of the throughput of a planetary nebula and discuss the effect of the bandpass shift in limiting the field of view that captures the wavelength of interest. We describe how utilizing DSLM as an integral field unit, rather than a simple narrowband imager, can mitigate the effects of the ultranarrow-bandpass and provide additional spectroscopic data utilizing the entire detector field of view.

\acknowledgments 

We are very grateful to the staff at New Mexico Skies Observatories, without whom this work couldn’t have been carried out. This work is supported by a Canadian Foundation for Innovation (CFI) grant. We are thankful for contributions from the Dunlap Institute (funded through an endowment established by the David Dunlap family and the University of Toronto), the Natural Sciences and Engineering Research Council of Canada (NSERC), and the National Science Foundation (NSF), without which this research would not have been possible. 
This research made use of Astropy,\footnote{http://www.astropy.org} a community-developed core Python package for Astronomy \cite{astropy:2013, astropy:2018}.

\bibliography{references_new} 
\bibliographystyle{spiebib} 

\end{document}